\documentclass[twocolumn,aps,amsmath,amssymb,prl,superscriptaddress,showpacs]{revtex4}
\usepackage{bm}
\usepackage{epsfig}
\usepackage[usenames]{color}
\usepackage{graphicx}

\usepackage[hypertex]{hyperref}

\DeclareMathOperator{\Tr}{Tr}

\renewcommand{\Re}{\mathop\mathrm{Re}\nolimits}
\renewcommand{\Im}{\mathop\mathrm{Im}\nolimits}

\renewcommand{\paragraph}[1]{\textit{#1.---} }

\begin{document}

\title{Synchronized Andreev Transmission in SNS Junction Arrays}
\author{N.\,M. Chtchelkatchev}
\affiliation{L.D.\ Landau Institute for
Theoretical Physics, Russian Academy of Sciences, 117940 Moscow,
Russia}
\affiliation{Materials Science Division, Argonne National Laboratory, Argonne, Illinois 60439, USA}
\affiliation{Department of Theoretical Physics, Moscow Institute
of Physics and Technology, 141700 Moscow, Russia}
\author{T.\,I. Baturina}
\affiliation{Materials Science Division, Argonne National Laboratory, Argonne, Illinois 60439, USA}
\affiliation{Institute of Semiconductor Physics, 13 Lavrentjev Avenue, Novosibirsk, 630090 Russia}
\author{A. Glatz}
\affiliation{Materials Science Division, Argonne National Laboratory, Argonne, Illinois 60439, USA}
\author{V.\,M. Vinokur}
\affiliation{Materials Science Division, Argonne National Laboratory, Argonne, Illinois 60439, USA}

\date{\today}

\begin{abstract}
We construct a nonequilibrium theory for the charge transfer through a diffusive
array of alternating normal (N)
and superconducting (S) islands comprising an SNSNS junction, with the size
of the central S-island being smaller than  the energy relaxation length.
We demonstrate that in the nonequilibrium regime the central island
acts as Andreev retransmitter with the Andreev conversions at both NS interfaces
of the central island correlated via over-the-gap transmission and Andreev reflection.
This results in a synchronized transmission at certain resonant voltages which can be
experimentally observed as a sequence of spikes in the differential conductivity.
\end{abstract}



\maketitle

An array of alternating superconductor (S) - normal metal (N) islands is a fundamental laboratory
representing a wealth of physical systems ranging from Josephson junction networks
and layered high temperature superconductors to disordered superconducting
films in the vicinity of the superconductor-insulator transition.
Electronic transport in these systems is mediated by Andreev conversion
of a supercurrent into a current of quasiparticles and vice versa at
interfaces between the superconducting and normal regions~\cite{Andreev}.
A fascinating phenomenon benchmarking this mechanism is the enhancement
of the conductivity observed in a single SNS junction at matching voltages constituting
an integer ($m$) fraction of the superconducting gap, $V=2\Delta/(em)$~\cite{SNSexp}
due to the effect of multiple Andreev reflection (MAR)~\cite{MAR}.
The current-voltage characteristics of diffusive
SNS junctions were studied in detail in~\cite{Shumeiko,Cuevas}.
Recent experimental findings~\cite{2DSNS,1DSNS,Fritz}
posed the question about the transport properties of large arrays comprised of many
SNS junctions.
\begin{figure}[b]
\begin{center}
\includegraphics[width=85mm]{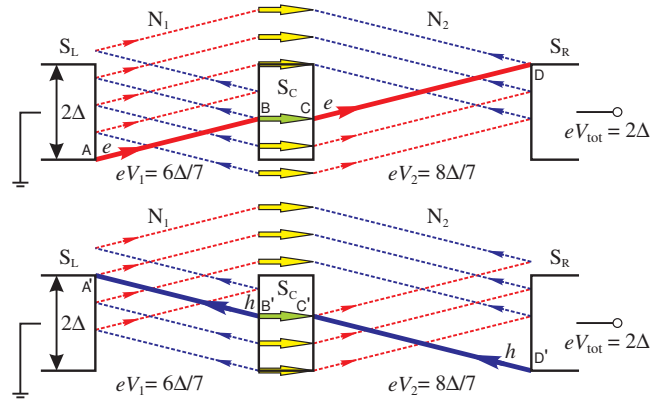}
\caption{(color online).
Diagrams of the SAT processes for $n=1$ in Eq.\,\eqref{eq:V_position}
in an SNSNS junction with normal resistance ratio $R_1/R_2=3/4$
(depicted through the 3/4 ratio of the respective lengths of the normal regions).
The thick solid lines represent the synchronized quasiparticle paths connecting
the points of singularity in the density of states corresponding to energies
$\varepsilon=\pm\Delta$ at the
electrodes S$_{_\mathrm L}$ and S$_{_\mathrm R}$.
The paths ABCD (upper panel)  and D$^\prime$C$^\prime$B$^\prime$A$^\prime$ (lower panel)
 correspond to the electron- and hole trajectories, respectively.
Synchronization of the energies of the incident and emitted quasiparticles
at points B and C (B$^\prime$ and C$^\prime$) is shown by arrows.
SAT is realized by trajectories passing through the singular points $\pm\Delta$
of the central island S$_{_\mathrm C}$ and including above-the-gap Andreev reflections.
Trajectories synchronizing other transmissions across S$_{_\mathrm C}$
and those of higher orders are not shown.
Note, that voltage drops $eV_1=6\Delta/7$ and $eV_2=8\Delta/7$ are not MAR matching voltages
of either the S$_{_\mathrm L}$N$_1$S$_{_\mathrm C}$
or the S$_{_\mathrm C}$N$_2$S$_{_\mathrm R}$ part.
}
\label{fig:fig1}
\end{center}
\end{figure}
In this Letter we develop a nonequilibrium theory for the current-voltage
characteristics of a series of two diffusive SNS junctions, i.e. an SNSNS junction.
The normal parts have, in general, different resistances and are
coupled via a small superconducting granule, S$_{_\mathrm C}$.
The focal point is the construction of a nonequilibrium circuit theory
for the charge transfer across the superconducting island.
We demonstrate that the wedging of S$_{_\mathrm C}$
into the normal part of an SNS junction
leads to nontrivial physics and the appearance of
a new distinct resonant mechanism for the current transfer:
{\it the Synchronized Andreev Transmission} (SAT).
In the SAT regime the processes of
Andreev conversion at the boundaries of the central superconducting island are correlated:
as a quasiparticle with energy $\varepsilon$ hits one
NS$_{_\mathrm C}$
interface, a quasiparticle with the same energy emerges from the other
S$_{_\mathrm C}$N interface
into the bulk of the normal island (and vice versa, see Fig.\,1).
This energy synchronization is achieved via over-the-gap
Andreev processes~\cite{1DSNS}, which couple the
MAR processes occurring in each of the normal islands and
make the quasiparticle distributions at the central island essentially in nonequilibrium.
Effectiveness of the synchronization is controlled by the
energy relaxation lengths of both, the quasiparticles crossing
S$_{_\mathrm C}$ with energies above $\Delta$, and of quasiparticles experiencing MAR in the
normal parts.
The SAT processes result in spikes in the differential conductivity of the
SNSNS circuit, which appear at resonant values of the \textit{total} applied voltage
$V_{\mathrm{tot}}$  defined by the condition
   \begin{gather}\label{eq:V_position}
       V_{\mathrm{tot}}= \frac{2\Delta}{en}
   \end{gather}
with integer $n$, irrespectively of the details of the distribution of the partial voltages
at the two normal islands.
As we show below, the SAT-induced features become dominant in large arrays consisting of
many SNS junctions.

\textit{The model.---}
We consider the charge transfer across
an S$_{_\mathrm L}$N$_1$S$_{_\mathrm C}$N$_2$S$_{_\mathrm R}$ junction, where
S$_{_\mathrm L}$, S$_{_\mathrm C}$, and S$_{_\mathrm R}$
are superconducting islands with identical gap $\Delta$.
We assume the size of the central island $L_{_\mathrm C}$ to be much larger than the
superconducting coherence length $\xi$, hence processes of subgap elastic
cotunneling and/or direct Andreev tunneling~\cite{Deutcher} are not effective.
In general this condition ensures that $L_{_\mathrm C}$
is large enough so that charges do not accumulate in the central island and
Coulomb blockade effects are irrelevant for the quasiparticle transport.
At the same time $L_{_\mathrm C}$ is assumed to be less than the
charge imbalance length, such that we can neglect the coordinate dependence of the
quasiparticle distribution functions across the island S$_{_\mathrm C}$.
Additionally, the condition $\ell_{\varepsilon}\gg L_{_\mathrm C}$,
where $\ell_{\varepsilon}$ is the energy relaxation length,
implies that quasiparticles with energies $\varepsilon>\Delta$ traverse the
central superconducting island S$_{_\mathrm C}$
without a loss of energy.
The normal parts N$_1$ and N$_2$ are the diffusive normal metals of length
$L_{1,2}>\xi$, and
$L_{1,2}>L_{_\mathrm T}$,
$L_{_\mathrm T}=\sqrt{\hbar D_{_\mathrm N}/\varepsilon}$,
where $D_{_\mathrm N}$ is the diffusion coefficient in the normal metal.
We assume the Thouless energy, $E_{\mathrm {Th}}=\hbar D_{_\mathrm N}/L_{1,2}^2$,
 to be small, $E_{\mathrm {Th}}\ll\Delta$,
and not to exceed the characteristic voltage drops,
$E_{\mathrm {Th}}<eV_{1,2}$.
These conditions are referred to as incoherent regime~\cite{Shumeiko}
where, in particular,
the Josephson coupling between the superconducting islands is suppressed.
We let the energy relaxation length in the normal parts N$_1$ and N$_2$
be much larger than their sizes, thus quasiparticles
may experience many incoherent Andreev reflections inside the normal regions.

The current transfer across the SNSNS junction is described by
quasiclassical Larkin-Ovchinnikov (LO) equations for the dirty limit~\cite{Larkin_Ovchinnikov}:
\begin{gather}\label{eq:LO}
 -i[\check{H}_{\rm eff}\circ,{\bf \check G}]=\nabla\mathbf{\check{J}},\quad \check {\mathbf {J}}\cdot \mathbf
 {n}=\frac{1}{2\sigma_{{_\mathrm S}}R}[\check{G}_{_\mathrm S}\,,\check{G}_{_\mathrm N}]\,,
\end{gather}
where $\check{H}_{\rm eff}=\check 1 (i\hat\sigma_z\partial_t-\varphi\hat\sigma_0+\hat{\Delta})$,
$\mathbf{\check{J}}=D{\bf \check G}\circ\nabla {\bf \check G}$
is the matrix current, the subscripts ``S" and ``N"
stand for superconducting and normal materials, respectively,
``$\circ$'' is the time-convolution, $\hat\sigma^\alpha$ ($\alpha=\{0,1,2,3\}$)
are the Pauli matrices,
$\hat\Delta=i\hat\sigma_{\mathrm x}\Im\Delta+i\hat\sigma_{\mathrm y}\Re\Delta$,
and $R$ is the resistance of an NS interface.
The diffusion coefficient $D$ assumes the value
$D_{_\mathrm N}$ in the normal metal and the value
$D_{_\mathrm S}$ in the superconductor, and $\varphi$
is the electrical potential which we calculate self-consistently.
The unit vector $\mathbf n$ is normal to the NS interface and is assumed
to be directed from N to S.
The momentum averaged Green's functions ${\bf \check G}({\bf r}, t, t^{\prime})$
are $2\times 2$ matrices in a Keldysh space.
Each element of the Keldysh matrix, labelled with a hat sign, is, in its turn,
a $2\times 2$ matrix in the electron-hole space:
       \begin{equation}
              \label{keldysh-space}
              {\bf \check G} = \left( \begin{array}{cc}
              \hat G^{\mathrm R} & \hat G^{\mathrm K} \\
               0     & \hat G^{\mathrm A}
             \end{array} \right); \hspace{3mm}
             \hat G^{{\mathrm R}({\mathrm A})} = \left( \begin{array}{cc}
              {\cal G}^{{\mathrm R}({\mathrm A})}
              & {\cal F}^{{\mathrm R}({\mathrm A})} \\
             \tilde {\cal F}^{{\mathrm R}({\mathrm A})}
             & \tilde {\cal G}^{{\mathrm R}({\mathrm A})}
          \end{array} \right)\,,
      \end{equation}
${\bf r}$ is the spatial position and $t$, $t^{\prime}$ are the two time arguments.
\noindent
The Keldysh component of the Green's function is parametrized as~\cite{Larkin_Ovchinnikov}:
$\hat G^{\mathrm K}=\hat G^{\mathrm R}\circ \hat{f}-\hat{f}\circ \hat G^{\mathrm A}$,
where $\hat f$ is the distribution function matrix, diagonal in Nambu space,
$\hat f\equiv \mathrm{ diag}\,[1-2n_{\mathrm e},1-2n_{\mathrm h}]$,
$n_{\mathrm {e(h)}}$ is the electron (hole) distribution function.
In equilibrium $n_{\mathrm {e(h)}}$ becomes the Fermi function.
And, finally, the Green's function satisfies the normalization
condition ${\bf \check G}^2=\check 1$.

The edge conditions closing Eqs.\,\eqref{eq:LO} are given by the expressions for the
Green's functions in the bulk of the left (L) and right (R) superconducting
leads: $${\bf \check G}_{{\mathrm L}({\mathrm R})}(t,t^{\prime}) =
e^{-i \mu_{{\mathrm L}({{\mathrm R})}}t \hat \tau_3/\hbar}
{\bf \check G}_0(t-t^{\prime}) e^{i \mu_{{\mathrm L}({{\mathrm R})}}t^\prime
\hat\tau_3/ \hbar}\,,$$
the chemical potentials are $\mu_{_\mathrm L}=0$ and $\mu_{_\mathrm R}=eV$.
Here, ${\bf \check G}_0(t)$ is the equilibrium bulk BCS Green's function.

The current density is expressed through the Keldysh component of $\mathbf{\check{J}}$ as
\begin{gather}\label{eq:IVnew}
\mathcal I(t,\mathbf r)=\frac {\pi\sigma_{_{\mathrm N}}}{4}\Tr \hat\sigma_z
\hat{J}^{\mathrm K}(t,t;\mathbf{r})=\frac {1}{2}\int d\varepsilon
\left[I_{\mathrm e}(\varepsilon)+I_{\mathrm h}(\varepsilon)\right]\,,
\end{gather}
where the spectral currents $I_{\mathrm e}$ and $I_{\mathrm h}$ are the time Wigner-transforms
of top- and bottom diagonal elements of the  matrix current
$\mathbf{\check J}^{({\mathrm K})}$,
representing electron and hole quasiparticle currents, respectively.
 In the bulk of a normal metal
$I_{\mathrm e}=\sigma_{_\mathrm N}\nabla n_{\mathrm e}$ and
$I_{\mathrm h}=\sigma_{_\mathrm N}\nabla n_{\mathrm h}$.

\begin{figure}[b]
\includegraphics[width=85mm]{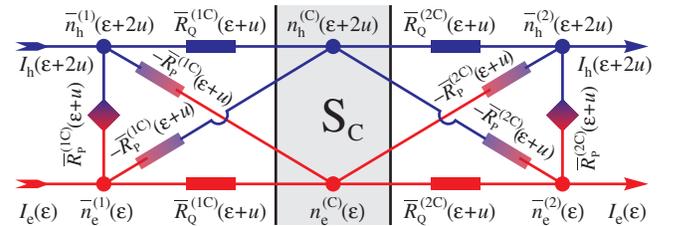}\\
\caption{(color online).
Effective circuit representing current conversion at the interfaces of
the central superconducting island S$_{_\mathrm C}$.
Resistors, $R_{_\mathrm{P}}$ and $R_{_\mathrm{Q}}$ stand for an Andreev- and a normal
processes, respectively.
 }\label{fig:boundary_eq}
\end{figure}

The distribution functions of quasiparticles at the central island S$_{_\mathrm C}$ are
essentially is nonequilibrium.  To take this into account we define
quasiparticle spectral currents at NS interfaces by the Keldysh component
of the boundary conditions for Eqs.~\eqref{eq:LO}.
These nonequilibrium boundary conditions have a form
of Kirchhoff's laws for the circuit shown in Fig.\,\ref{fig:boundary_eq}.
The electron and hole distribution functions take the role of voltages at the nodes.
For illustration we write down the equation for an electronic spectral current flowing into the
lower left corner node (Kirchhoff's laws at the other corner nodes have a similar form):
\begin{multline}\label{eq:Ie}
I_{\mathrm e}(\varepsilon)=\frac{1}{\bar{R}_{_\mathrm Q}^{(\scriptscriptstyle{\mathrm{1C}})}(\varepsilon+u)}
\left[n_{\mathrm e}^{{\scriptscriptstyle({\mathrm C})}}(\varepsilon)-
\bar{n}_{\mathrm e}^{({1})}(\varepsilon)\right]-
\\
\frac{1}{\bar{R}_{_\mathrm P}^{(\scriptscriptstyle{\mathrm{1C}})}(\varepsilon+u)}\,\left[ {n}_{\mathrm h}^{{\scriptscriptstyle{(\mathrm C})}}
(\varepsilon+2u)- \bar{n}_{\mathrm e}^{({1})}(\varepsilon)\right]+
\\
\frac{1}{\bar{R}_{_\mathrm P}^{(\scriptscriptstyle{\mathrm{1C}})}(\varepsilon+u)}\,\left[ \bar{n}_{\mathrm h}^{({1})}
(\varepsilon+2u)- \bar{n}_{\mathrm e}^{({1})}(\varepsilon)\right]
\,.
\end{multline}
The interjacent resistances, $\bar{R}_{{_\mathrm Q({_\mathrm P})}}$, are
  defined as $\bar{R}_{{_\mathrm Q({_\mathrm P})}}^{-1}(\varepsilon)=
  \{\bar{R}^{-1}_{-}(\varepsilon)\pm \bar{R}^{-1}_{+}(\varepsilon)\}/2$ where
   $\bar{R}_{\pm}(\varepsilon)$ are the special functions characterizing transparencies
 of the interfaces  and tabulated in~\cite{Shumeiko}.
   At high energies, $|\varepsilon|\gg\Delta$,
 $\bar R_{+}(\varepsilon)\to \bar R_{-}(\varepsilon)$; at small energies, $|\varepsilon|\ll\Delta$, $\bar R_{-}(\varepsilon)$
diverges, while $\bar R_{+}(\varepsilon)-\bar R_{-}(\varepsilon)$ remains finite; and there are singularities
in $\partial_\varepsilon\bar R_{\pm}(\varepsilon)$ at $|\epsilon|=\Delta$, which are of the same origin
as those in the BCS density of states and reflect the fact that
 quasiparticles cannot penetrate the superconductor below the gap.
The ``bars'' indicate that the respective resistances and the distribution
functions are renormalized by the proximity effect.

\begin{figure}[b]
\includegraphics[width=85mm]{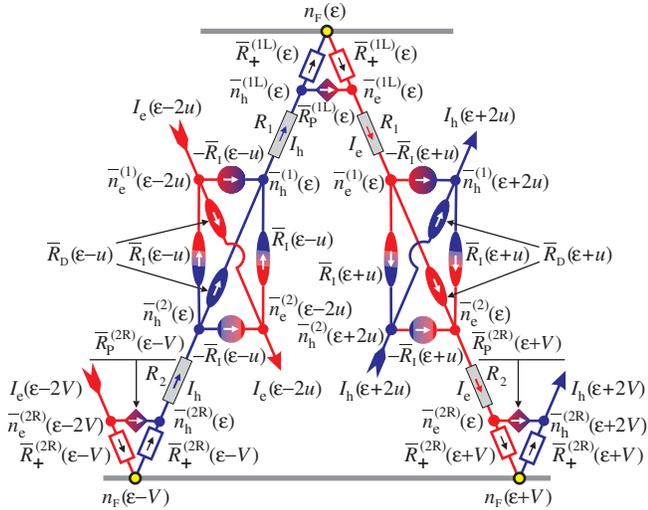}\\
\caption{(color online).
Effective circuit for an SNSNS junction.
To simplify the graph, resistors $\bar R_{{_\mathrm D({_\mathrm I})}}^{-1}
=(1/2)\{
 [\bar R_{+}^{({{\scriptscriptstyle{\mathrm {1C}}}})}
+\bar R_{+}^{({{\scriptscriptstyle{\mathrm {2C}}}})}]^{-1}
\pm [\bar R_{-}^{({{\scriptscriptstyle{\mathrm {1C}}}})}
+\bar R_{-}^{({{\scriptscriptstyle{\mathrm {2C}}}})}]^{-1}\}$ representing
Andreev and normal processes at the NS interfaces of the central island are introduced.
}\label{fig:eff-circuit}
\end{figure}

To derive the current-voltage characteristics
for the general case of an asymmetric nonequilibrium SNSNS junction with
different resistances of the normal regions, we construct a nonequilibrium circuit
theory allowing for an analytical solution of the nonlinear nonuniform matrix Eqs.~\eqref{eq:LO}
for Keldysh-Nambu Green's functions.  The diagrammatic mapping of Eqs.\,\eqref{eq:LO} is realized
by an equivalent circuit
shown in Fig.\,\ref{fig:eff-circuit}.
The Kirchhoff's equations for the potential distribution in the circuit
of Fig.\,\ref{fig:eff-circuit}, give the recurrent relations:

\begin{gather}\label{eq:recurrent1}
\begin{split}
\mathcal{R}(\varepsilon,-{\mathrm u},- V)
I_{\mathrm h}(\varepsilon)-\rho^{(\circ)}({\varepsilon-{\mathrm u}})
I_{\mathrm e}(\varepsilon-2{\mathrm u})\\
-\rho^{(\triangleright)}(\varepsilon)I_{\mathrm e}(\varepsilon)
-\rho^{(\triangleleft)}(\varepsilon -V)I_{\mathrm e}(\varepsilon- V)
\\
=n_{\scriptscriptstyle{\mathrm F}}(\varepsilon)-n_{\scriptscriptstyle{\mathrm F}}(\varepsilon-V),
\end{split}
\\ \label{eq:recurrent2}
\begin{split}
\mathcal{R}(\varepsilon,{\mathrm u}, V)I_{\mathrm e}(\varepsilon)
- \rho^{(\circ)}(\varepsilon+{\mathrm u})I_{\mathrm h}(\varepsilon+2{\mathrm u}) \\
- \rho^{(\triangleright)}(\varepsilon) I_{\mathrm h}(\varepsilon)
- \rho^{(\triangleleft)}(\varepsilon+ V)I_{\mathrm h}(\varepsilon+2V)
\\
=n_{\scriptscriptstyle{\mathrm F}}(\varepsilon+V)-n_{\scriptscriptstyle{\mathrm F}}(\varepsilon)\,,
\end{split}
\end{gather}
where the electric potential $u$ of the
S$_{\scriptscriptstyle{\mathrm C}}$ island is calculated  self-consistently from the
electroneutrality condition, $u=(\pi/8)\Tr \hat G^{\scriptscriptstyle{\mathrm K}}$.
The effective resistance is
$\mathcal{R}=R_{{1}}+R_{2}+\rho^{(\triangleright\circ\triangleleft)}$, where
$\rho^{(\triangleright\circ\triangleleft)}=
(1/2)\sum_{\alpha=\pm}
\{\bar R_{\alpha,\varepsilon}^{({{\scriptscriptstyle{\mathrm {1L}}}})}+
  \bar R_{\alpha,\varepsilon+{\mathrm u}}^{({{\scriptscriptstyle{\mathrm {1C}}}})}+
  \bar R_{\alpha,\varepsilon+{\mathrm u}}^{({{\scriptscriptstyle{\mathrm {2C}}}})}+
  \bar R_{\alpha,\varepsilon+\scriptscriptstyle{\mathrm V}}^{({{\scriptscriptstyle{\mathrm {2R}}}})}\}$,
$\rho^{(\circ)}=(1/2)\{
 \bar R_{+}^{({{\scriptscriptstyle{\mathrm {1C}}}})}
+\bar R_{+}^{({{\scriptscriptstyle{\mathrm {2C}}}})}
-\bar R_{-}^{({{\scriptscriptstyle{\mathrm {1C}}}})}
-\bar R_{-}^{({{\scriptscriptstyle{\mathrm {2C}}}})}\}$,
$\rho^{(\triangleleft)}=(1/2)\{
 \bar R_{+}^{({{\scriptscriptstyle{\mathrm {2R}}}})}
-\bar R_{-}^{({{\scriptscriptstyle{\mathrm {2R}}}})}\}$,
and
$\rho^{(\triangleright)}=(1/2)\{
 \bar R_{+}^{({{\scriptscriptstyle{\mathrm {1L}}}})}
-\bar R_{-}^{({{\scriptscriptstyle{\mathrm {1L}}}})}\}$.
Solutions of Eqs.\eqref{eq:recurrent1} yield the required $I$-$V$ characteristics
for an asymmetric SNSNS junction.  To verify our formulas we note that
at large quasiparticle energies, $|\varepsilon|\gg\Delta$, the total resistance $\mathcal{R}$
reduces to the normal resistance of the array, whereas $\rho^{(\triangleleft)}$ and
$\rho^{(\triangleright)}$ vanish.
Then we find from Eqs.\eqref{eq:recurrent1}-\eqref{eq:recurrent2}
that
$I_{\mathrm h}(\varepsilon)=
[n_{\scriptscriptstyle{\mathrm F}}(\varepsilon)-n_{\scriptscriptstyle{\mathrm F}}(\varepsilon-V)]
/\mathcal{R}$,
and
$I_{\mathrm e}(\varepsilon)=
[n_{\scriptscriptstyle{\mathrm F}}(\varepsilon+V)-n_{\scriptscriptstyle{\mathrm F}}(\varepsilon)]
/\mathcal{R}$,
which together with Eq.\eqref{eq:IVnew}
reproduce Ohm's law, $\mathcal I=V/\mathcal{R}$.
The constructed diagram, Fig.~\ref{fig:eff-circuit}, is an elemental building unit
for a general nonequilibrium quantitative theory of SNS arrays comprised of many SNS junctions.

\begin{figure}[t]
\begin{center}
\includegraphics[width=70mm]{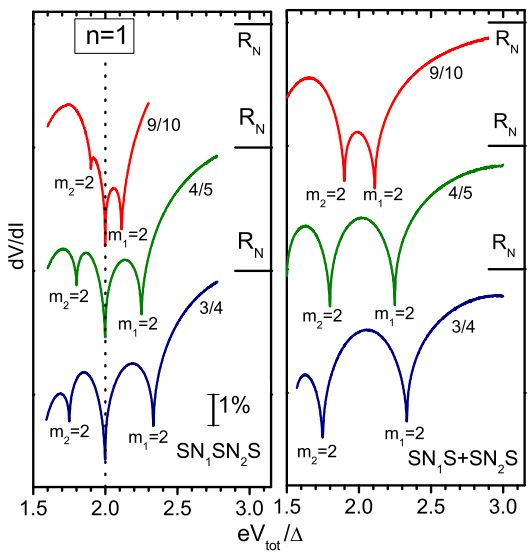}
\caption{(color online). Left panel: Differential resistances as functions of the applied
voltage $V_{\mathrm{tot}}$ (around $n=1$ in Eq.\,(1))
for the SN$_1$SN$_2$S junction.
The fractions 3/4, and 4/5, and 9/10
represent the ratios of resistances of the normal regions, $R_1/R_2$.
$dV/dI$ of SN$_1$SN$_2$S junction demonstrates
the pronounced SAT spike at $V_{\mathrm{tot}}=2\Delta/e$,
irrespectively of the partial voltage drops.
The SAT spike is sandwiched between the two additional spikes corresponding to
individual MAR processes occurring at junctions SN$_1$S and SN$_2$S for $m_{1}, m_{2}=2$.
The voltage positions of these features
depends on $R_1/R_2$.
Right panel:
The corresponding $dV/dI(V_1+V_2)$ for two SN$_1$S and SN$_2$S junctions in series
as they would have appeared in absence of the synchronization process,
i.e. when $L_{\scriptscriptstyle{\mathrm C}} > \ell_{\varepsilon}$. These $dV/dI$
were calculated following~\cite{Shumeiko} (with transmissivity W=1).
}
\label{fig:fig4}
\end{center}
\end{figure}

\textit{Results and discussion.---} The calculation of the current-voltage characteristics ${\mathcal I}(V)$
requires the numerical solution of the recurrent relations, Eqs.~\eqref{eq:recurrent1}-\eqref{eq:recurrent2}.
To this end, we have developed a computational scheme allowing to bypass instabilities
caused by the non-analytic behavior of the spectral currents $I_{{\mathrm e}({\mathrm h})}(\varepsilon)$.
We first fix some chosen energy $\varepsilon$, identify the set of energies connected through
the equations in the given energy interval, and solve the resulting subsystem of equations.
We then repeat the procedure, until
the required energy resolution of $\delta\varepsilon=10^{-5}\Delta$ is achieved.
Typically, up to $10^6$ linear equations had to be solved for every given voltage, but the complexity of the coupled subsystem depends on the commensurability of $u$ and $V$.

Figure~\ref{fig:fig4} shows the comparative results for the SNSNS junction and two SNS junctions in series.
The latter corresponds to the case where the size of the central island well exceeds
the energy relaxation length, $L_{\scriptscriptstyle{\mathrm C}} > \ell_{\varepsilon}$.
We display the differential resistances as functions of the applied
voltage, which demonstrate the singularities in
Andreev transmission more profoundly than the $I$-$V$ curves.
There is a pronounced SAT spike in the $dV/dI$ for an SNSNS junction at $V_{\mathrm{tot}}=2\Delta/e$.
The spike appears irrespectively of the partial voltage drops in the normal regions
and is absent in the corresponding curves representing two individual MAR processes
at the junctions SN$_1$S and SN$_2$S.

The resonant voltages of the SAT singularities can be found from the  consideration of the quasiparticle
trajectories in the space-energy diagrams.  Such a diagram for the first subharmonic, $n=1$ and ratio
$R_1/R_2=3/4$ is given in Fig.\,\ref{fig:fig1}.  A quasiparticle starts from the left superconducting electrode with
the energy $\varepsilon=-\Delta$ to traverse N$_1$, and
the quasiparticle that starts from the central island S$_{\mathrm c}$
with the same energy as the incident one to take up upon the current across the
island N$_2$, and hit S$_{\scriptscriptstyle{\mathrm R}}$ with
the energy $\varepsilon=\Delta$ (the ABCD path, the corresponding path
for the hole is D$^\prime$C$^\prime$B$^\prime$A$^\prime$).
In general, relevant trajectories yielding resonant voltages of Eq.\,\eqref{eq:V_position}
have the following structure:  they start and end at the BCS quasiparticle density of states singular
points ($\varepsilon=\pm\Delta$), contain the closed polygonal path, which include
MAR staircases in the normal parts and over-the-gap transmissions and Andreev reflections,
and pass the density of states singular points at the central island.
Apart from the main singularities
[Eq.~\eqref{eq:V_position}], additional SAT satellite spikes
appear at $V=(2\Delta/e)(p+q)/n$, where $p/q$ is the irreducible rational approximation
of the real number $r=R_1/R_2$,
(we take $R_1<R_2$), and $n\geqslant(p+q)$.

The achieved qualitative understanding enables us to observe
that the manifestations
of the SAT mechanism in an experimental situation
becomes even more pronounced with the growth of the number of SNS junctions in the
system.  To see this, let us assume that the resistances of the normal islands in a chain
of SNS junctions are randomly scattered around their
average value $R_0$ and follow Gaussian statistics with the standard deviation
$\sigma_{\scriptscriptstyle{\mathrm R}}=\sigma R_0$, where $\sigma$ is dimensionless.
Accordingly, the dispersion of the distribution of the MAR resonant voltages is characterized by
the same $\sigma$, and the MAR features get smeared.
Let us distribute the voltage drop
$2\Delta/e$ among the $n$ successive islands.
Then the quasiparticle SAT path starts at the lower edge of
the superconducting gap at island $j$, traverses $n-1$ intermediate superconducting islands
and hits the edge of the gap at the $j+n$-th island in the chain.
The standard deviation of the voltage drop on the $n$ islands grows
as $\sqrt{n}$ resulting in a voltage deviation per one island $\propto 1/\sqrt{n}$,
i.e. the dispersion of the distribution of $V_{\mathrm n}$ drops
with increasing $n$: $\sigma_{\scriptscriptstyle{\mathrm{SAT}}}=\sigma/\sqrt{n}$.
In contrast to the MAR-induced features, with an increase of $n$, the subharmonic spikes at voltages
$V_{\mathrm n}$ per junction due to SAT processes become more sharp and pronounced.

In conclusion, we have developed a nonequilibrium theory of charge transfer across a
superconducting island and found that the island acts as Andreev {\it retransmitter}.
We have shown that the nonequilibrium transport through an SNSNS array is governed
by synchronized Andreev transmission with correlated conversion processes
at the NS interfaces.
The constructed theory is a fundamental building unit for a general quantitative
description of a large  array consisting of many SNS junctions.

\paragraph{Acknowledgements}
We thank A.\,N.\,Omelyanchuk for helpful discussions.
The work was supported by the
U.S. Department of Energy Office of Science under the Contract No. DE-AC02-06CH11357,
by the Russian Foundation for Basic
Research (Grants Nos. 10-02-00700 and 09-02-01205), the Dynasty, and the Russian Academy of Science Programs.

\end{document}